\documentclass[aps,preprint,nofootinbib]{revtex4-1}
\usepackage{amssymb,amsmath,mathtools}
\usepackage{graphicx}
\usepackage{subcaption}
\usepackage[format=plain,labelsep=period,justification=justified,singlelinecheck=false]{caption}
\usepackage{placeins}
\graphicspath{{figs/}}
\newcommand{\tilelayoutscale}{0.945}
\newcommand{\tilelayout}[1]{\includegraphics[scale=\tilelayoutscale]{#1}}
\usepackage{ragged2e}
\usepackage{xcolor}
\usepackage{tikz}
\usetikzlibrary{arrows.meta,positioning,calc}
\usepackage[colorlinks=true,citecolor=blue,linkcolor=blue]{hyperref}

\newcommand{\A}{\mathsf{A}}
\newcommand{\B}{\mathsf{B}}
\newcommand{\C}{\mathsf{C}}
\newcommand{\D}{\mathsf{D}}
\newcommand{\E}{\mathsf{E}}
\newcommand{\F}{\mathsf{F}}
\newcommand{\G}{\mathsf{G}}
\renewcommand{\H}{\mathsf{H}}

\newcommand{\CX}{\mathsf{CX}}
\newcommand{\MX}{\mathsf{MX}}
\newcommand{\MZ}{\mathsf{MZ}}
\newcommand{\RX}{\mathsf{RX}}
\newcommand{\RZ}{\mathsf{RZ}}

\begin{document}

\begin{center}
{\large\bfseries Block algebra for morphing circuits\par}
Rui Chao\\
NVIDIA
\end{center}

\begin{center}
\begin{minipage}{0.92\textwidth}
\noindent\textbf{Abstract.}
Morphing circuits are a new paradigm for quantum error correction that relaxes hardware requirements.
We present four constructions for CNOT-based CSS morphing circuits with explicit qubit connectivity degrees.
All four constructions are specified in block algebra notation, with entries in algebras generated by permutation matrices.
The first three are obtained by rewriting existing surface- and color-code morphing circuits;
the fourth is a new three-round construction modeled on the 6.6.6 color code.
The surface-code construction recovers the morphing circuit of Ref.~\cite{shaw2025lowering} for two-block group algebra codes.
Numerical search then instantiates these permutation matrices using regular representations of finite groups.
\end{minipage}
\end{center}

\section{Introduction}\label{sec:introduction}

Morphing circuits are syndrome-extraction circuits in which the physical qubits of a stabilizer code
also serve as the temporary measurement workspace.
They relax hardware requirements by replacing extra measurement ancillas with repeated contractions of the code itself.
Each contraction round applies Clifford gates that turn selected stabilizer checks into single-qubit
Paulis, measures and resets the corresponding qubits, and then reverses the contraction.
This organization appears in the hex-grid surface-code circuit~\cite{mcewen2023relaxing},
the middle-out color-code circuits~\cite{gidney2023new}, and the general morphing-circuit
framework~\cite{shaw2025lowering}.

This paper develops a block-algebra description of these circuits.
We apply the standard algebraic viewpoint for translation-invariant
codes~\cite{haah2016algebraic,liang2025generalized,steffan2025tile} to morphing circuits.
We group qubits and checks into blocks of equal size, so each block entry of $H_X$ or $H_Z$
is a permutation matrix or a formal sum of such matrices over $\mathbb F_2$.
In this notation, a CNOT layer acts as a coupled column operation on $H_X$ and $H_Z$.
A contraction schedule is therefore a Gaussian-elimination pattern over block stabilizer matrices.
The repeated translation structure in known surface- and color-code circuits becomes an algebraic
template: selected translation monomials can be promoted to polynomial sums, subject only to the
commutation constraints needed for CSS orthogonality and valid contractions.

Table~\ref{tab:constructions} summarizes the four resulting constructions.
Construction~I is based on the hex-grid surface-code circuit of Ref.~\cite{mcewen2023relaxing}
and recovers the morphing circuit of Ref.~\cite{shaw2025lowering} for two-block group algebra codes.
Constructions~II and III are based on the middle-out color-code circuits of Ref.~\cite{gidney2023new}.
Construction~IV is a three-round construction modeled on the 6.6.6 color code.
This block form facilitates numerical searches over finite groups.
The rest of the paper is organized as follows.
Section~\ref{sec:framework} fixes the block notation and the stabilizer evolution rule.
Sections~\ref{sec:cI}--\ref{sec:cIV} give the four constructions, including the contraction
schedules, the measured qubit blocks after contraction, and the end-cycle stabilizer matrices.
Section~\ref{sec:numerics} gives the finite-group code search and the circuit-level simulations.

\begin{table}[!htbp]
\caption{\justifying Four constructions for morphing circuits, listing the stabilizer matrices
of the CSS mid-cycle code, constraints on the parameters, and qubit connectivity degree.
Lowercase letters $p,q,r$ denote permutation matrices, and uppercase letters $P,Q,R,S$ denote
formal sums of such matrices, with $|\cdot|$ denoting the number of summands.
In the stabilizer matrices, the symbol $1$ denotes the identity matrix, and overlines denote
matrix transposition.}
\label{tab:constructions}
\centering
\small
\setlength{\arraycolsep}{2pt}
\begin{tabular}{@{}c@{\quad}|@{\quad}r@{\,}c@{\,}l@{\quad}c@{\quad}c@{}}
\hline\hline
& \multicolumn{3}{c}{Mid-cycle code} & Constraints & Connectivity \\
\hline
I & $\begin{bmatrix} H_X \\\hline H_Z \end{bmatrix}$ & $=$ &
$\begin{bmatrix}
1 & P & Q & 1 \\
R & 1 & 1 & S \\\hline
1 & \overline R & \overline S & 1 \\
\overline P & 1 & 1 & \overline Q
\end{bmatrix}$ &
$\begin{array}{c}
PR=QS\\
RP=SQ
\end{array}$ &
$\begin{array}{c}
\max\{|P|+|R|,\\[-1pt]
|Q|+|S|\}+1
\end{array}$ \\
II & $H_X=H_Z$ & $=$ &
$\begin{bmatrix}
P & P & 1 & q \\
\overline q & 1 & \overline P & \overline P
\end{bmatrix}$ &
$Pq=qP$ &
$|P|+1$ \\
III & $H_X=H_Z$ & $=$ &
$\begin{bmatrix}
1 & 1 & \cdot & \cdot & 1 & 1 & \cdot & \cdot \\
\cdot & \cdot & 1 & 1 & \cdot & \cdot & 1 & 1 \\
P & P & 1 & r & Q & Q & 1 & r \\
\overline r & 1 & \overline P & \overline P & \overline r & 1 & \overline Q & \overline Q
\end{bmatrix}$ &
$\begin{array}{c}
Pr=rP\\
Qr=rQ
\end{array}$ &
$\max\{|P|,|Q|\}+2$ \\
IV & $H_X=H_Z$ & $=$ &
$\begin{bmatrix}
1 & 1 & 1 & p & p & p \\
q & q & 1 & 1 & 1 & p \\
\overline p & 1 & 1 & 1 & \overline q & \overline q
\end{bmatrix}$ &
$pq=qp$ &
$3$ \\
\hline\hline
\end{tabular}
\end{table}

\clearpage
\section{Common framework}\label{sec:framework}

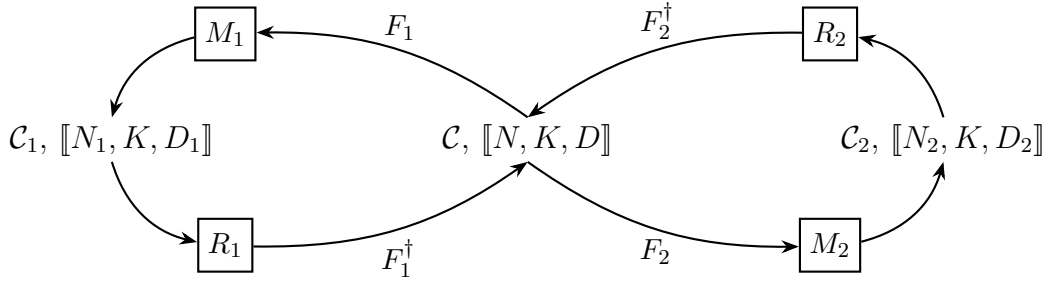
\begin{figure}[t]
\begin{center}
\begin{tikzpicture}[
    >=Stealth,
    thick,
    box/.style={draw, rectangle, minimum width=0.7cm, minimum height=0.7cm, font=\small},
    code/.style={font=\normalsize, inner sep=2pt}]
  \node[code] (c1) at (-5.5, 0) {$\mathcal C_1,\,[\![N_1,K,D_1]\!]$};
  \node[code] (c)  at (0, 0)    {$\mathcal C,\,[\![N,K,D]\!]$};
  \node[code] (c2) at (5.5, 0)  {$\mathcal C_2,\,[\![N_2,K,D_2]\!]$};
  \node[box] (m1) at (-4, 1.4)  {$M_1$};
  \node[box] (r2) at (4, 1.4)   {$R_2$};
  \node[box] (r1) at (-4, -1.4) {$R_1$};
  \node[box] (m2) at (4, -1.4)  {$M_2$};
  \draw[->] (c.north) to[bend right=18] node[above, pos=0.5]{$F_1$} (m1);
  \draw[->] (m1) to[bend right=30] (c1.north);
  \draw[->] (c2.north) to[bend right=30] (r2);
  \draw[->] (r2) to[bend right=18] node[above, pos=0.5]{$F_2^\dagger$} (c.north);
  \draw[->] (c1.south) to[bend right=30] (r1);
  \draw[->] (r1) to[bend right=18] node[below, pos=0.5]{$F_1^\dagger$} (c.south);
  \draw[->] (c.south) to[bend right=18] node[below, pos=0.5]{$F_2$} (m2);
  \draw[->] (m2) to[bend right=30] (c2.south);
\end{tikzpicture}
\end{center}
\caption{\justifying A morphing circuit with $J=2$ contraction rounds; reproduced from
Ref.~\cite[Fig.~1]{shaw2025lowering}. The mid-cycle code $\mathcal C$ with $[\![N,K,D]\!]$
sits at the center; the two end-cycle codes $\mathcal C_1, \mathcal C_2$ with
$[\![N_j,K,D_j]\!]$ and $N_j<N$ sit at the lobes. Starting from $\mathcal C_1$, the cycle
applies $R_1, F_1^\dagger$ to reach $\mathcal C$, then $F_2, M_2$ to reach $\mathcal C_2$;
it returns to $\mathcal C_1$ via $R_2, F_2^\dagger$ and $F_1, M_1$.}
\label{fig:bowtie}
\end{figure}

\paragraph{Morphing circuit.}
Given a stabilizer code $\mathcal C$ with $[\![N,K,D]\!]$, a \emph{morphing
circuit}~\cite{shaw2025lowering} is a syndrome-extraction circuit where each operation -- gate,
measurement, reset -- acts only on the $N$ physical qubits of $\mathcal C$.
A cycle of syndrome extraction consists of a sequence of $J$ \emph{contraction rounds}.
Specifically, round $j\in[J]$ applies gates $F_j$ that contract a subset of check operators of
$\mathcal C$ into single-qubit Paulis, measures their values, resets the measured qubits, and
applies $F_j^\dagger$ to return to $\mathcal C$.
The unmeasured qubits in round $j$ support the \emph{end-cycle code} $\mathcal C_j$ with
$[\![N_j,K,D_j]\!]$; $\mathcal C$ is called the \emph{mid-cycle code}.
See Fig.~\ref{fig:bowtie}.

\paragraph{Permutation matrix.}
The $N$ physical qubits of $\mathcal C$ are partitioned into \emph{blocks} of equal size $n$,
labeled with sans-serif capitals $\A,\B,\C,\ldots$.
Lowercase letters $p, q, \ldots$ denote monomials, represented by $n\times n$ permutation matrices
over $\mathbb F_2$.
Uppercase letters $P, Q, \ldots$ denote polynomials, formal sums of such monomials.
For $P=\sum_{\alpha=1}^{|P|}p_\alpha$, we write $|P|$ for the number of monomial terms.
Matrix transposition is denoted by an overline: $\overline P:=P^\top$, $\overline p_\alpha:=p_\alpha^\top$.

Circuit operations are restricted to $\CX$ gates, single-qubit measurements $\MX, \MZ$, and
single-qubit resets $\RX, \RZ$. For example, $\MX(\A)$ measures every qubit of $\A$ in the
$X$-basis; $\RZ(\B)$ resets every qubit of $\B$ to $|0\rangle$; $\CX(\A, p, \B)$ applies
$n$ parallel $\CX$ gates from $\A$ to $\B$, paired by a permutation matrix $p$ -- rows of $p$
index $\A$, columns index $\B$ (qubit $i$ of $\A$ controls qubit $j$ of $\B$ if and only if
$p_{ij}=1$). For a polynomial $P=\sum_{\alpha=1}^{|P|}p_\alpha$, the shorthand
$\CX(\A, P, \B)$ denotes the $|P|$ layers $\CX(\A, p_\alpha, \B)$, performed in any order.

\paragraph{Translation invariance.}
Our principal examples of the mid-cycle code are surface and color codes on a 2D torus, whose
physical qubits and check operators are translation-invariant. Two integer $2\times 2$ matrices
with row vectors specify the geometry: $A$ with $\det A =: N$ defines a lattice whose cosets
$\mathbb Z^2/\mathbb Z^2 A$ index the $N$ physical qubits in the torus; and $B$ with
$\mathbb Z^2 A \subseteq \mathbb Z^2 B$ defines a superlattice whose cosets
$\mathbb Z^2/\mathbb Z^2 B$ index the qubits in each \emph{site}. The torus then contains
$\det A/\det B =: n$ sites indexed by $\mathbb Z^2 B/\mathbb Z^2 A$.

\begin{figure}[t]
\centering
\begin{minipage}[c]{0.32\linewidth}\centering
  \tilelayout{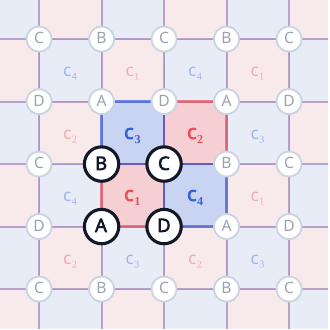}
\end{minipage}\hspace{6pt}%
\begin{minipage}[c]{0.10\linewidth}\raggedright
  $x=(2,0)$\\[3pt]$y=(0,2)$
\end{minipage}\hspace{6pt}%
\begin{minipage}[c]{0.50\linewidth}\centering
  \setlength\tabcolsep{6pt}
  $\begin{array}{c|cccc}
        & \A & \B & \C & \D \\\hline
   \mathsf{c}_1  & 1  & 1  & 1  & 1  \\
   \mathsf{c}_2  & xy & x  & 1  & y  \\\hline
   \mathsf{c}_3  & y  & 1  & 1  & y  \\
   \mathsf{c}_4  & x  & x  & 1  & 1  \\[-1pt]
   \multicolumn{5}{c}{\rule{0pt}{2.2ex}}
  \end{array} = \begin{bmatrix} H_X \\\hline H_Z \end{bmatrix}$
\end{minipage}
\caption{\justifying Surface code: site layout (left) with translation vectors $x, y$, and
parity-check matrices (right). Plaquettes $\mathsf{c}_1, \mathsf{c}_2$ support $X$-checks;
$\mathsf{c}_3, \mathsf{c}_4$ support $Z$-checks.}
\label{fig:surface}
\end{figure}

Figs.~\ref{fig:surface}, \ref{fig:666_2round}, \ref{fig:488}, and~\ref{fig:tw666} show the
four principal examples; each figure pairs the site layout --- specified by the rows $x, y$ of
$B$ --- with the parity-check matrices of the mid-cycle code. Each site (highlighted in the layout)
carries qubits labeled $\A, \B, \C, \ldots$ and plaquettes labeled
$\mathsf{c}_1, \mathsf{c}_2, \ldots$. A qubit block is the set of $n$ translates of a single
qubit position across the $n$ sites; plaquettes are grouped into blocks the same way. The
parity-check matrices thus have rows indexed by plaquette blocks and columns indexed by qubit
blocks. Each entry is a polynomial acting on the $n$ sites: $1$ is the identity, and $x$ and $y$
are the commuting monomials induced by the corresponding site translations. When both $X$- and
$Z$-stabilizer matrices are displayed, we stack them vertically with a separating hline. Blank
entries denote zeros.

\paragraph{Stabilizer evolution rule.}
Let $(P_X, Q_X)$ and $(P_Z, Q_Z)$ be block rows of $X$- and $Z$-type stabilizers supported on
$\A\sqcup\B$, where each component is a polynomial. The layer $\CX(\A, p, \B)$ maps
\begin{equation}\label{eq:cxrule}
(P_X, Q_X) \mapsto (P_X,\ Q_X + P_X\,p)\enspace ,\qquad (P_Z, Q_Z) \mapsto (P_Z + Q_Z\,\overline p,\ Q_Z)\enspace .
\end{equation}
Outside the translation-invariant examples, monomials need not commute; products such as $P_X p$
and $Q_Z\overline p$ are ordered as displayed.

Each contraction $F_j$ starts from the stabilizer matrices $H_X$ and $H_Z$ of $\mathcal C$ and
can be read as Gaussian elimination on their block rows. By Eq.~\eqref{eq:cxrule}, each layer
$\CX(\A, p, \B)$ in $F_j$ implements a coupled column operation: in the current $X$-stabilizer
matrix, the $\A$-column right-multiplied by $p$ is added to the $\B$-column; in the current
$Z$-stabilizer matrix, the $\B$-column right-multiplied by $\overline p$ is added to the
$\A$-column. We write
\[
\mathcal E_X^j := F_j H_X F_j^\dagger\enspace ,\qquad
\mathcal E_Z^j := F_j H_Z F_j^\dagger
\]
for the stabilizer matrices obtained by evolving $H_X$ and $H_Z$ through $F_j$ according to these
updates. For $\mathcal E=\mathcal E_X^j$ or $\mathcal E_Z^j$, $\left.\mathcal E\right|_\A$
denotes the evolved stabilizer matrix restricted to the $\A$-column. The layers in $F_j$ are
chosen so that selected stabilizer rows evolve to single-column monomial pivots, whose
corresponding qubit blocks are then measured out. The surviving rows and columns form the
stabilizer matrices $H_X^j$ and $H_Z^j$ of $\mathcal C_j$.

The following equivalence operations on parity-check matrices are useful. Left-multiplying a block
row of $H_X$ or $H_Z$ by a monomial relabels checks; right-multiplying a block column by a monomial,
simultaneously in $H_X$ and $H_Z$, relabels qubits~\cite[Theorem~6]{lin2024quantum}. These
operations put the concrete surface- and color-code matrices into template forms whose monomial
entries can readily be promoted to polynomials in the general constructions below.

\section{Construction~I}\label{sec:cI}

\begin{figure}[t]
\centering

\begin{subfigure}{\linewidth}\centering
\begin{minipage}[t]{0.37\linewidth}\vspace*{0pt}\centering
  $\begin{array}{c|cccc}
        & \A & \B & \C & \D \\\hline
   \mathsf{c}_1  & 1  & \textcolor{blue!50}{1}  & \textcolor{blue!50}{1}  & 1  \\
   \mathsf{c}_2  & \textcolor{blue!50}{xy} & x  & 1  & \textcolor{blue!50}{y}  \\\hline
   \mathsf{c}_3  & y  & \textcolor{blue!50}{1}  & \textcolor{blue!50}{1}  & y  \\
   \mathsf{c}_4  & \textcolor{blue!50}{x}  & x  & 1  & \textcolor{blue!50}{1}  \\[-1pt]
   \multicolumn{5}{c}{\rule{0pt}{2.2ex}}
  \end{array} = \begin{bmatrix} H_X \\\hline H_Z \end{bmatrix}$
\end{minipage}\hfill
\begin{minipage}[t]{0.60\linewidth}\vspace*{0pt}\centering
  \setlength\tabcolsep{2pt}
  \begin{tabular}{l|ll|ll}
            & \multicolumn{2}{c|}{$F_1$} & \multicolumn{2}{c}{$F_2$} \\\hline
    Stage 1 & $\CX(\A, 1, \B)$       & $\CX(\D, 1, \C)$           & $\CX(\B, y, \A)$ & $\CX(\C, y, \D)$ \\
    Stage 2 & $\CX(\D, 1, \A)$       & $\CX(\C, x, \B)$           & $\CX(\A, 1, \D)$ & $\CX(\B, \overline x, \C)$ \\
    Measure & $\MX(\D)$ & $\MZ(\B)$ & $\MX(\B)$ & $\MZ(\D)$
  \end{tabular}
\end{minipage}
\vspace{-1cm}
\subcaption{}
\label{fig:surface-schedule-surface}
\end{subfigure}

\begin{subfigure}{\linewidth}\centering
\vspace{2mm}
\begin{minipage}[t]{0.37\linewidth}\vspace*{0pt}\centering
  $\begin{array}{c|cccc}
        & \A & \B & \C & \D \\\hline
   \mathsf{c}_1  & 1  & P  & Q  & 1  \\
   \mathsf{c}_2  & R  & 1  & 1  & S  \\\hline
   \mathsf{c}_3  & 1  & \overline R  & \overline S  & 1  \\
   \mathsf{c}_4  & \overline P  & 1  & 1  & \overline Q  \\[-1pt]
   \multicolumn{5}{c}{\rule{0pt}{2.2ex}}
  \end{array} = \begin{bmatrix} H_X \\\hline H_Z \end{bmatrix}$
\end{minipage}\hfill
\begin{minipage}[t]{0.60\linewidth}\vspace*{0pt}\centering
  \setlength\tabcolsep{2pt}
  \begin{tabular}{l|ll|ll}
            & \multicolumn{2}{c|}{$F_1$} & \multicolumn{2}{c}{$F_2$} \\\hline
    Stage 1 & $\CX(\A, P, \B)$ & $\CX(\D, Q, \C)$ & $\CX(\B, R, \A)$ & $\CX(\C, S, \D)$ \\
    Stage 2 & $\CX(\D, 1, \A)$ & $\CX(\C, 1, \B)$ & $\CX(\A, 1, \D)$ & $\CX(\B, 1, \C)$ \\
    Measure & $\MX(\D)$ & $\MZ(\B)$ & $\MX(\B)$ & $\MZ(\D)$
  \end{tabular}
\end{minipage}
\vspace{-1cm}
\subcaption{}
\label{fig:surface-schedule-cI}
\end{subfigure}
\vspace{-1cm}
\caption{\justifying (a)~Mid-cycle stabilizers (left) and contractions (right) for the surface
code, which reproduce the hex-grid circuit of Ref.~\cite[Fig.~8]{mcewen2023relaxing}. Blue entries
in the stabilizer matrices are eliminated by the Stage~1 layers of $F_1$ and $F_2$. (b)~Mid-cycle
stabilizers (left) and contractions (right) for Construction~I, which reproduce the two-block
group algebra morphing circuit of Ref.~\cite[Table~III]{shaw2025lowering}. The polynomial entries
satisfy $PR = QS$ and $RP = SQ$.}
\label{fig:surface-schedule}
\end{figure}

Fig.~\ref{fig:surface-schedule-surface} shows the hex-grid surface-code circuit of
Ref.~\cite{mcewen2023relaxing}, and Fig.~\ref{fig:surface-schedule-cI} shows Construction~I,
which reproduces the two-block group algebra morphing circuit of Ref.~\cite{shaw2025lowering}.
Both the mid-cycle stabilizers and contractions in Fig.~\ref{fig:surface-schedule-cI} mirror those
in Fig.~\ref{fig:surface-schedule-surface}, but with selected monomials replaced by polynomials.
In Fig.~\ref{fig:surface-schedule-surface}, the Stage~1 layers of $F_1$ and $F_2$ eliminate the
blue entries. As equivalence operations of Sec.~\ref{sec:framework}, right-multiplying the
$\B$-column by $\overline x$ and left-multiplying the $\mathsf{c}_3$-row by $\overline y$ put the
surface-code matrices into the template of Fig.~\ref{fig:surface-schedule-cI}, with
$P=\overline x$, $Q=1$, $R=xy$, and $S=y$. In the general construction, $P,Q,R,S$ are arbitrary
polynomials subject only to $PR=QS$ and $RP=SQ$, enforced by CSS orthogonality. The qubit
connectivity degree is at most $1+\max\{|P|+|R|,\, |Q|+|S|\}$.

Every stabilizer row of the mid-cycle code is contracted under $F_1$ or $F_2$ to a single-column monomial pivot, shown in blue:
\begin{equation}\label{eq:cI-evolved}
\begin{array}{@{}c@{\quad}c@{\quad}c@{\quad}c@{}}
\left.\mathcal E_X^1\right|_\D =
\begin{bmatrix} \textcolor{blue!50}{1} \\ S \end{bmatrix}
&
\left.\mathcal E_Z^1\right|_\B =
\begin{bmatrix} \overline R \\ \textcolor{blue!50}{1} \end{bmatrix}
&
\left.\mathcal E_X^2\right|_\B =
\begin{bmatrix} P \\ \textcolor{blue!50}{1} \end{bmatrix}
&
\left.\mathcal E_Z^2\right|_\D =
\begin{bmatrix} \textcolor{blue!50}{1} \\ \overline Q \end{bmatrix}
\end{array}\enspace .
\end{equation}
After measuring $\B\sqcup\D$, the surviving rows and columns form $\mathcal C_1$ and $\mathcal C_2$ on $\A\sqcup\C$,
\begin{equation}\label{eq:cI-end}
\begin{bmatrix} H_X^1 \\\hline H_Z^1 \end{bmatrix} =
\begin{bmatrix} R+S & 1+SQ \\\hline 1+\overline R\,\overline P & \overline R+\overline S \end{bmatrix}\enspace ,
\qquad
\begin{bmatrix} H_X^2 \\\hline H_Z^2 \end{bmatrix} =
\begin{bmatrix} 1+PR & P+Q \\\hline \overline P+\overline Q & 1+\overline Q\,\overline S \end{bmatrix}\enspace .
\end{equation}

Construction~I recovers Ref.~\cite[Table~III]{shaw2025lowering} through the homomorphism criterion
of Ref.~\cite[Criterion~C.4]{shaw2025lowering}. The homomorphism $f:G\to\mathbb Z_2$ splits each
of the left ($L$) and right ($R$) qubit blocks into $K=\ker f$ and $K^c=G\setminus K$ sectors;
after ordering these sectors as $L_K,R_{K^c},L_{K^c},R_K$, the Table~III contractions have the
form of Fig.~\ref{fig:surface-schedule-cI}. In this embedding, $Q$ and $R$ are the two coset
restrictions of the action of $A\setminus\{a_1\}$, and likewise $P$ and $S$ for
$B\setminus\{b_1\}$. Thus Ref.~\cite{shaw2025lowering} implicitly imposes $|Q|=|R|$ and
$|P|=|S|$. Construction~I keeps the same block form but imposes only CSS orthogonality, so these
weights may differ.

\section{Construction~II}\label{sec:cII}

\begin{figure}[t]
\centering
\begin{minipage}[c]{0.32\linewidth}\centering
  \tilelayout{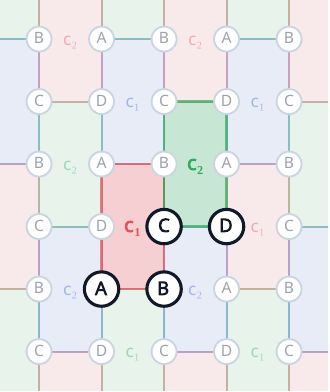}
\end{minipage}\hspace{6pt}%
\begin{minipage}[c]{0.10\linewidth}\raggedright
  $x=(2,0)$\\[3pt]$y=(0,2)$
\end{minipage}\hspace{6pt}%
\begin{minipage}[c]{0.50\linewidth}\centering
  \setlength\tabcolsep{6pt}
  $\begin{array}{c|cccc}
        & \A     & \B     & \C     & \D     \\\hline
   \mathsf{c}_1  & 1{+}y  & 1{+}y  & 1      & \overline x \\
   \mathsf{c}_2  & xy     & y      & 1{+}y  & 1{+}y \\[-1pt]
   \multicolumn{5}{c}{\rule{0pt}{2.2ex}}
  \end{array} = H_X = H_Z$
\end{minipage}
\caption{\justifying 6.6.6 color code: site layout (left) with translation vectors $x, y$,
and parity-check matrices (right). Each plaquette supports both an $X$- and a $Z$-check.}
\label{fig:666_2round}
\end{figure}

\begin{figure}[t]
\centering

\begin{subfigure}{\linewidth}\centering
\begin{minipage}[t]{0.40\linewidth}\vspace*{0pt}\centering
  $\begin{array}{c|cccc}
        & \A & \B & \C & \D \\\hline
   \mathsf{c}_1  & \textcolor{blue!50}{1{+}y}  & \textcolor{blue!50}{1{+}y}  & 1            & \overline x \\
   \mathsf{c}_2  & xy     & y      & \textcolor{blue!50}{1{+}y}        & \textcolor{blue!50}{1{+}y} \\[-1pt]
   \multicolumn{5}{c}{\rule{0pt}{2.2ex}}
  \end{array} = H_X = H_Z$
\end{minipage}\hspace{8pt}%
\begin{minipage}[t]{0.55\linewidth}\vspace*{0pt}\centering
  \setlength\tabcolsep{2pt}
  \begin{tabular}{l|ll}
            & \multicolumn{2}{c}{$F_1$} \\\hline
    Stage 1 & $\CX(\C, 1{+}y, \B)$ & $\CX(\D, x(1{+}y), \A)$ \\
    Stage 2 & $\CX(\C, \overline x, \D)$ & $\CX(\A, \overline x, \B)$ \\
    Measure & $\MX(\C)$ & $\MZ(\B)$
  \end{tabular}
\end{minipage}
\subcaption{}
\label{fig:cII-schedule-color}
\end{subfigure}

\begin{subfigure}{\linewidth}\centering
\begin{minipage}[t]{0.40\linewidth}\vspace*{0pt}\centering
  $\begin{array}{c|cccc}
        & \A & \B & \C & \D \\\hline
   \mathsf{c}_1  & P            & P            & 1            & q \\
   \mathsf{c}_2  & \overline q  & 1            & \overline P  & \overline P \\[-1pt]
   \multicolumn{5}{c}{\rule{0pt}{2.2ex}}
  \end{array} = H_X = H_Z$
\end{minipage}\hspace{8pt}%
\begin{minipage}[t]{0.55\linewidth}\vspace*{0pt}\centering
  \setlength\tabcolsep{2pt}
  \begin{tabular}{l|ll}
            & \multicolumn{2}{c}{$F_1$} \\\hline
    Stage 1 & $\CX(\C, P, \B)$ & $\CX(\D, \overline q P, \A)$ \\
    Stage 2 & $\CX(\C, q, \D)$ & $\CX(\A, q, \B)$ \\
    Measure & $\MX(\C)$ & $\MZ(\B)$
  \end{tabular}
\end{minipage}
\subcaption{}
\label{fig:cII-schedule-template}
\end{subfigure}
\vspace{-1cm}
\caption{\justifying Two-round morphing circuits for (a)~the 6.6.6 color code and
(b)~Construction~II. For each circuit, $F_2$ is the Hadamard conjugate of $F_1$ and is not shown.
Blue entries in the color-code stabilizers are eliminated by the Stage~1 layers of $F_1$ and
$F_2$. The mid-cycle stabilizers of Construction~II are subject to $Pq = qP$.}
\label{fig:cII-schedule}
\end{figure}

Fig.~\ref{fig:cII-schedule-color} shows a two-round morphing circuit for the 6.6.6 color code.
Since the mid-cycle code is self-dual, $F_2$ can be chosen as the Hadamard conjugate of the
displayed $F_1$, and thus is not shown.
The Stage~1 layers of $F_1$ and $F_2$ eliminate the blue translation polynomials in the stabilizer matrix.
Construction~II promotes these polynomials to arbitrary polynomials.
Specifically, right-multiplying the $\A$- and $\B$-columns of the color-code stabilizer matrix by
$\overline y$ puts it into the template of Fig.~\ref{fig:cII-schedule-template}, with
$P=1+\overline y$ and $q=\overline x$.
In the general construction, the polynomial $P$ and monomial $q$ are arbitrary, subject only to
$Pq=qP$, enforced by CSS orthogonality.
The qubit connectivity degree is $1+|P|$.

Every stabilizer row of the mid-cycle code is contracted under $F_1$ or $F_2$ to a single-column monomial pivot, shown in blue:
\begin{equation}\label{eq:cII-evolved}
\begin{array}{@{}c@{\qquad}c@{}}
\left.\mathcal E_X^1\right|_\C =
\left.\mathcal E_Z^2\right|_\C =
\begin{bmatrix} \textcolor{blue!50}{1} \\ \overline P \end{bmatrix}
&
\left.\mathcal E_Z^1\right|_\B =
\left.\mathcal E_X^2\right|_\B =
\begin{bmatrix} P \\ \textcolor{blue!50}{1} \end{bmatrix}
\end{array}\enspace .
\end{equation}
After measuring $\B\sqcup\C$, the surviving rows and columns form $\mathcal C_1$ and $\mathcal C_2$ on $\A\sqcup\D$,
\begin{equation}\label{eq:cII-end}
\begin{bmatrix} H_X^1 \\\hline H_Z^1 \end{bmatrix} =
\begin{bmatrix} H_Z^2 \\\hline H_X^2 \end{bmatrix} =
\begin{bmatrix} \overline q(1+\overline P P) & (1+q)\overline P \\\hline (1+\overline q)P & q(1+P\overline P) \end{bmatrix}\enspace .
\end{equation}

Fig.~\ref{fig:cII-schedule-color} is based on the middle-out circuit of
Ref.~\cite[Fig.~4]{gidney2023new} for the 6.6.6 color code, but uses a cleaner pivot choice.
That circuit corresponds to a variant in which Stage~1 first eliminates a polynomial except for
one constituent monomial, then uses this remaining monomial as the pivot. Choose a constituent
monomial $p$ of $P$ such that $pq=qp$, and write $P=p+P'$. The corresponding variant of the $F_1$
in Fig.~\ref{fig:cII-schedule-template} is
\begin{equation}\label{eq:cII-pivot-variant}
\begin{array}{c|ll}
& \multicolumn{2}{c}{F_1'} \\\hline
\text{Stage }1a & \CX(\C,P',\B) & \CX(\D,\overline q P',\A) \\
\text{Stage }1b & \CX(\B,\overline p,\C) & \CX(\A,\overline p q,\D) \\
\text{Stage }2 & \CX(\B,1,\A) & \CX(\D,1,\C) \\
\text{Measure} & \MX(\B) & \MZ(\C)
\end{array}\enspace .
\end{equation}
After measuring $\B\sqcup\C$, the surviving rows and columns form $\mathcal C_1'$ and $\mathcal C_2'$ on $\A\sqcup\D$,
\begin{equation}\label{eq:cII-pivot-end}
\begin{bmatrix} H_X^{1,\prime} \\\hline H_Z^{1,\prime} \end{bmatrix} =
\begin{bmatrix} H_Z^{2,\prime} \\\hline H_X^{2,\prime} \end{bmatrix} =
\begin{bmatrix}
(1+\overline q)(1+\overline P P') & (1+\overline P P)\overline p \\\hline
(1+P\overline P)p & (1+q)(1+P\overline{P'})
\end{bmatrix}\enspace .
\end{equation}
This variant has the same qubit connectivity degree $1+|P|$.
The same pivot variant applies to the other constructions: it switches the relevant $\CX$
directions and changes which qubits and bases are measured, but does not change the qubit
connectivity degree.

\section{Construction~III}\label{sec:cIII}

\begin{figure}[t]
\centering
\begin{minipage}[c]{0.32\linewidth}\centering
  \tilelayout{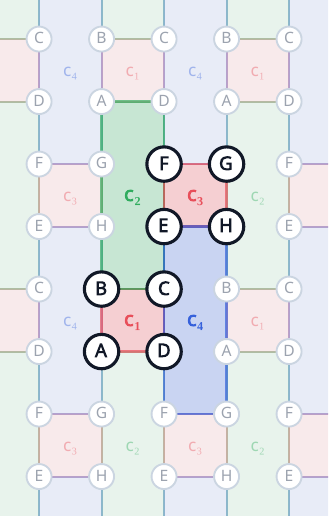}
\end{minipage}\hspace{6pt}%
\begin{minipage}[c]{0.10\linewidth}\raggedright
  $x=(2,0)$\\[3pt]$y=(0,4)$
\end{minipage}\hspace{6pt}%
\begin{minipage}[c]{0.54\linewidth}\centering
  \setlength\tabcolsep{4pt}
  $\begin{array}{c|cccccccc}
        & \A & \B & \C & \D & \E & \F & \G & \H \\\hline
   \mathsf{c}_1  & 1  & 1  & 1  & 1  &    &    &    &    \\
   \mathsf{c}_2  & y  & 1  & 1  & y  & 1  & 1  & \overline x & \overline x \\
   \mathsf{c}_3  &    &    &    &    & 1  & 1  & 1  & 1  \\
   \mathsf{c}_4  & x  & x  & 1  & 1  & 1  & \overline y & \overline y & 1 \\[-1pt]
   \multicolumn{9}{c}{\rule{0pt}{2.2ex}}
  \end{array} = H_X = H_Z$
\end{minipage}
\caption{\justifying 4.8.8 color code: site layout (left) with translation vectors $x, y$,
and parity-check matrices (right). Each plaquette supports both an $X$- and a $Z$-check.}
\label{fig:488}
\end{figure}

\begin{figure}[tp]
\centering

\begin{subfigure}[t]{0.49\linewidth}\centering
\begin{minipage}[t]{\linewidth}\vspace*{0pt}\centering
  \setlength\tabcolsep{1.2pt}
  $\begin{array}{c|cccccccc}
        & \A & \B & \C & \D & \E & \F & \G & \H \\\hline
   \mathsf{c}_1  & 1  & 1  & 1  & 1  &    &    &    &    \\
   \mathsf{c}_2  & \textcolor{blue!50}{y}  & \textcolor{blue!50}{1}  & \textcolor{blue!50}{1}  & \textcolor{blue!50}{y}  & 1  & 1  & \overline x & \overline x \\
   \mathsf{c}_3  &    &    &    &    & 1  & 1  & 1  & 1  \\
   \mathsf{c}_4  & x  & x  & 1  & 1  & \textcolor{blue!50}{1}  & \textcolor{blue!50}{\overline y}  & \textcolor{blue!50}{\overline y} & \textcolor{blue!50}{1} \\[-1pt]
   \multicolumn{9}{c}{\rule{0pt}{2.2ex}}
  \end{array} = H_X = H_Z$
\end{minipage}
\vspace{1mm}
\begin{minipage}[t]{\linewidth}\vspace*{0pt}\centering
  \setlength\tabcolsep{1pt}
  \begin{tabular}{@{}l|ll@{}}
            & \multicolumn{2}{c}{$F_1$} \\\hline
    Stage 1 & $\CX(\E, 1, \C)$ & $\CX(\F, y, \D)$ \\
            & $\CX(\G, xy, \A)$ & $\CX(\H, x, \B)$ \\
    Stage 2 & $\CX(\B, 1, \A)$ & $\CX(\C, 1, \D)$ \\
            & $\CX(\F, 1, \E)$ & $\CX(\G, 1, \H)$ \\
    Stage 3 & $\CX(\D, x, \A)$ & $\CX(\G, x, \F)$ \\
            & $\CX(\E, 1, \H)$ & $\CX(\B, 1, \C)$ \\
    Measure & $\MX(\B)$ & $\MX(\G)$ \\
            & $\MZ(\A)$ & $\MZ(\H)$
  \end{tabular}
\end{minipage}
\subcaption{}
\label{fig:cIII-schedule-color}
\end{subfigure}\hfill%
\begin{subfigure}[t]{0.49\linewidth}\centering
\begin{minipage}[t]{\linewidth}\vspace*{0pt}\centering
  \setlength\tabcolsep{1.2pt}
  $\begin{array}{c|cccccccc}
        & \A & \B & \C & \D & \E & \F & \G & \H \\\hline
   \mathsf{c}_1  & 1  & 1  & 1  & 1  &    &    &    &    \\
   \mathsf{c}_2  & P  & Q  & Q  & P  & 1  & 1  & r & r \\
   \mathsf{c}_3  &    &    &    &    & 1  & 1  & 1  & 1  \\
   \mathsf{c}_4  & \overline r  & \overline r  & 1  & 1  & \overline Q & \overline P & \overline P & \overline Q \\[-1pt]
   \multicolumn{9}{c}{\rule{0pt}{2.2ex}}
  \end{array} = H_X = H_Z$
\end{minipage}
\vspace{1mm}
\begin{minipage}[t]{\linewidth}\vspace*{0pt}\centering
  \setlength\tabcolsep{1pt}
  \begin{tabular}{@{}l|ll@{}}
            & \multicolumn{2}{c}{$F_1$} \\\hline
    Stage 1 & $\CX(\E, Q, \C)$ & $\CX(\F, P, \D)$ \\
            & $\CX(\G, \overline r P, \A)$ & $\CX(\H, \overline r Q, \B)$ \\
    Stage 2 & $\CX(\B, 1, \A)$ & $\CX(\C, 1, \D)$ \\
            & $\CX(\F, 1, \E)$ & $\CX(\G, 1, \H)$ \\
    Stage 3 & $\CX(\D, \overline r, \A)$ & $\CX(\G, \overline r, \F)$ \\
            & $\CX(\E, 1, \H)$ & $\CX(\B, 1, \C)$ \\
    Measure & $\MX(\B)$ & $\MX(\G)$ \\
            & $\MZ(\A)$ & $\MZ(\H)$
  \end{tabular}
\end{minipage}
\subcaption{}
\label{fig:cIII-schedule-template}
\end{subfigure}

\begin{subfigure}{\linewidth}\centering
\vspace{1mm}
  \setlength\tabcolsep{2pt}
  \renewcommand{\arraystretch}{0.9}
  $\begin{array}{c|cccccccc}
        & \A & \D & \F & \G & \B & \C & \E & \H \\\hline
   \mathsf{c}_1  & 1 & 1 &   &   & 1 & 1 &   &   \\
   \mathsf{c}_3  &   &   & 1 & 1 &   &   & 1 & 1 \\
   \mathsf{c}_2  & P & P & 1 & r & Q & Q & 1 & r \\
   \mathsf{c}_4  & \overline r & 1 & \overline P & \overline P
                & \overline r & 1 & \overline Q & \overline Q
  \end{array}\,\raisebox{-2.0ex}{$= H_X = H_Z$}$
  \renewcommand{\arraystretch}{1}
\subcaption{}
\label{fig:cIII-schedule-permuted}
\end{subfigure}
\vspace{-1cm}
\caption{\justifying Two-round morphing circuits for (a)~the 4.8.8 color code, reproducing the
middle-out circuit of Ref.~\cite{gidney2023new}, and (b)~Construction~III. For each circuit,
$F_2$ is the Hadamard conjugate of $F_1$ and is not shown. Blue entries in the color-code
stabilizers are eliminated by the Stage~1 layers of $F_1$ and $F_2$. The mid-cycle stabilizers
of Construction~III are subject to $Pr = rP$ and $Qr = rQ$. Panel~(c) shows the mid-cycle
stabilizers of Construction~III with reordered rows and columns.}
\label{fig:cIII-schedule}
\end{figure}

Fig.~\ref{fig:cIII-schedule-color} shows a two-round morphing circuit for the 4.8.8 color code,
which reproduces the middle-out circuit of Ref.~\cite{gidney2023new}.
For this circuit, $F_2$ is the Hadamard conjugate of the displayed $F_1$ and is not shown.
The Stage~1 layers of $F_1$ and $F_2$ eliminate the blue translation monomials in the color-code stabilizers.
Construction~III is obtained in Fig.~\ref{fig:cIII-schedule-template} by promoting these monomials
$y$ and $1$ to polynomials $P$ and $Q$, and replacing the notation $\overline x$ by a monomial $r$.
The validity of Construction~III requires $Pr=rP$ and $Qr=rQ$; apart from these commutation constraints, the parameters are arbitrary.
The qubit connectivity degree is $2+\max\{|P|,|Q|\}$.
With rows and columns reordered as in Fig.~\ref{fig:cIII-schedule-permuted}, the mid-cycle
stabilizers display two copies of the Construction~II stabilizers in
Fig.~\ref{fig:cII-schedule-template}, with parameters $(P,r)$ and $(Q,r)$, coupled by the
low-weight rows $\mathsf{c}_1$ and $\mathsf{c}_3$.

Every stabilizer row of the mid-cycle code is contracted under $F_1$ or $F_2$ to a single-column monomial pivot, shown in blue:
\begin{equation}\label{eq:cIII-evolved}
\begin{array}{@{}r@{\;}c@{\;}r@{\;}c@{\;}l@{\qquad}r@{\;}c@{\;}r@{\;}c@{\;}l@{}}
\left.\mathcal E_X^1\right|_\B
&=& \left.\mathcal E_Z^2\right|_\B
&=&
\begin{bmatrix} \textcolor{blue!50}{1} \\ \phantom{0} \\ \overline r Q \\ \overline r(1+\overline Q Q) \end{bmatrix}
&
\left.\mathcal E_X^1\right|_\G
&=& \left.\mathcal E_Z^2\right|_\G
&=&
\begin{bmatrix} \phantom{0} \\ \textcolor{blue!50}{r} \\ 1 \\ \overline P \end{bmatrix}
\\[1ex]
\left.\mathcal E_Z^1\right|_\H
&=& \left.\mathcal E_X^2\right|_\H
&=&
\begin{bmatrix} r\overline Q \\ r(1+Q\overline Q) \\ \textcolor{blue!50}{1} \\ \phantom{0} \end{bmatrix}
&
\left.\mathcal E_Z^1\right|_\A
&=& \left.\mathcal E_X^2\right|_\A
&=&
\begin{bmatrix} 1 \\ P \\ \phantom{0} \\ \textcolor{blue!50}{\overline r} \end{bmatrix}
\end{array}\enspace .
\end{equation}
After measuring $\A\sqcup\B\sqcup\G\sqcup\H$, the surviving rows and columns form $\mathcal C_1$ and $\mathcal C_2$ on $\C\sqcup\D\sqcup\E\sqcup\F$,
\begin{equation}\label{eq:cIII-end}
\begin{bmatrix} H_X^1 \\\hline H_Z^1 \end{bmatrix}
=
\begin{bmatrix} H_Z^2 \\\hline H_X^2 \end{bmatrix}
=
\begin{bmatrix}
(1+\overline r)Q                       & P+Q                         &                         & 1+\overline r \\
(1+\overline r)(1+\overline Q Q)       & \overline P P+\overline Q Q & \overline P+\overline Q & (1+\overline r)\overline P \\\hline
                                        & 1+r                         & (1+r)\overline Q        & \overline P+\overline Q \\
P+Q                                     & (1+r)P                      & (1+r)(1+Q\overline Q)  & P\overline P+Q\overline Q
\end{bmatrix}\enspace .
\end{equation}

\FloatBarrier
\section{Construction~IV}\label{sec:cIV}

\begin{figure}[!htbp]
\centering
\begin{minipage}[c]{0.381\linewidth}\centering
  \tilelayout{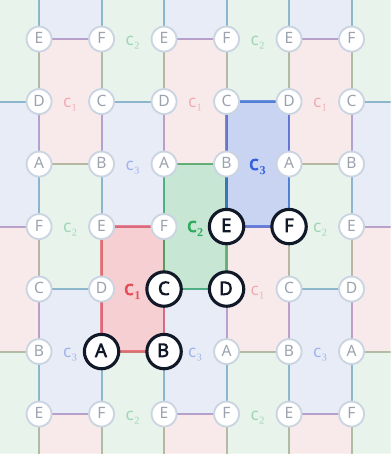}
\end{minipage}\hspace{6pt}%
\begin{minipage}[c]{0.10\linewidth}\raggedright
  $x=(2,0)$\\[3pt]$y=(1,3)$
\end{minipage}\hspace{6pt}%
\begin{minipage}[c]{0.40\linewidth}\centering
  \setlength\tabcolsep{4pt}
  $\begin{array}{c|cccccc}
        & \A & \B & \C & \D & \E & \F \\\hline
   \mathsf{c}_1  & 1  & 1  & 1  & \overline x & \overline x & \overline x \\
   \mathsf{c}_2  & y  & y  & 1  & 1  & 1  & \overline x \\
   \mathsf{c}_3  & xy & y  & y  & y  & 1  & 1 \\[-1pt]
   \multicolumn{7}{c}{\rule{0pt}{2.2ex}}
  \end{array}\,\raisebox{-0.5ex}{$= H_X = H_Z$}$
\end{minipage}
\caption{\justifying 6.6.6 color code with six qubits per site: site layout (left) with
translation vectors $x, y$, and parity-check matrices (right). Each plaquette supports both an
$X$- and a $Z$-check.}
\label{fig:tw666}

\vspace{5mm}

\setlength\tabcolsep{4pt}
$\begin{array}{c|cccccc}
        & \A & \B & \C & \D & \E & \F \\\hline
   \mathsf{c}_1  & 1  & 1  & 1  & p & p & p \\
   \mathsf{c}_2  & q  & q  & 1  & 1 & 1 & p \\
   \mathsf{c}_3  & \overline p & 1 & 1 & 1 & \overline q & \overline q \\[-1pt]
   \multicolumn{7}{c}{\rule{0pt}{2.2ex}}
  \end{array}\,\raisebox{-0.5ex}{$= H_X = H_Z$}$

\vspace{-3mm}

\setlength\tabcolsep{4pt}
\begin{tabular}{c|ll|ll|ll}
          & \multicolumn{2}{c|}{$F_1$} & \multicolumn{2}{c|}{$F_2$} & \multicolumn{2}{c}{$F_3$} \\\hline
  Stage 1 &
    $\CX(\E, q, \B)$ & $\CX(\F, \overline p q, \A)$ &
    $\CX(\A, p, \D)$ & $\CX(\B, 1, \C)$ &
    $\CX(\C, p, \F)$ & $\CX(\D, 1, \E)$ \\
  Stage 2 &
    $\CX(\C, p, \F)$ & $\CX(\D, 1, \E)$ &
    $\CX(\E, q, \B)$ & $\CX(\F, \overline p q, \A)$ &
    $\CX(\A, p, \D)$ & $\CX(\B, 1, \C)$ \\
  Stage 3 &
    $\CX(\B, 1, \C)$ & $\CX(\A, p, \D)$ &
    $\CX(\D, 1, \E)$ & $\CX(\C, p, \F)$ &
    $\CX(\F, \overline p q, \A)$ & $\CX(\E, q, \B)$ \\
  Stage 4 &
    \multicolumn{2}{l|}{$\CX(\C, 1, \D)$} &
    \multicolumn{2}{l|}{$\CX(\E, 1, \F)$} &
    \multicolumn{2}{l}{$\CX(\A, 1, \B)$} \\
  Measure &
    $\MX(\C)$ & $\MZ(\D)$ &
    $\MX(\E)$ & $\MZ(\F)$ &
    $\MX(\A)$ & $\MZ(\B)$
\end{tabular}
\caption{\justifying Three-round morphing circuit for Construction~IV. The mid-cycle stabilizers
(top) are subject to $pq=qp$. The contractions (bottom) respectively measure $X$-checks on
$\mathsf{c}_2,\mathsf{c}_3,\mathsf{c}_1$ and $Z$-checks on
$\mathsf{c}_3,\mathsf{c}_1,\mathsf{c}_2$ in rounds $F_1,F_2,F_3$.}
\label{fig:cIV-schedule}
\end{figure}

Construction~IV in Fig.~\ref{fig:cIV-schedule} has $J=3$ contraction rounds and is modeled on the
stabilizer matrices of the 6.6.6 color code with six qubits per site in Fig.~\ref{fig:tw666}.
Specifically, left-multiplying the $\mathsf{c}_3$-row by $\overline y$ and replacing the notation
$\overline x$ by a monomial $p$ and $y$ by a monomial $q$ put the mid-cycle stabilizers into the
displayed form.
The monomials $p$ and $q$ are arbitrary subject only to $pq=qp$.
Round $F_j$ measures the $X$-checks on $\mathsf{c}_{j+1}$ and the $Z$-checks on
$\mathsf{c}_{j+2}$, with indices taken modulo $3$; the measured qubit blocks advance by
$\sigma=(\A\,\C\,\E)(\B\,\D\,\F)$ from one round to the next.
The qubit connectivity degree is $3$.

\begin{samepage}
Every stabilizer row of the mid-cycle code is contracted under $F_1$, $F_2$, or $F_3$ to a single-column monomial pivot, shown in blue:
\begin{equation}\label{eq:cIV-evolved}
\begin{array}{@{}r@{}l@{\quad}r@{}l@{\quad}r@{}l@{\quad}r@{}l@{\quad}r@{}l@{\quad}r@{}l@{}}
\left.\mathcal E_X^1\right|_\C
&{}=
\begin{bmatrix} pq \\ \textcolor{blue!50}{1} \\ 1 \end{bmatrix}
&
\left.\mathcal E_Z^1\right|_\D
&{}=
\begin{bmatrix} \overline q \\ 1 \\ \textcolor{blue!50}{1} \end{bmatrix}
&
\left.\mathcal E_X^2\right|_\E
&{}=
\begin{bmatrix} p \\ pq \\ \textcolor{blue!50}{\overline q} \end{bmatrix}
&
\left.\mathcal E_Z^2\right|_\F
&{}=
\begin{bmatrix} \textcolor{blue!50}{p} \\ \overline q \\ \overline q \end{bmatrix}
&
\left.\mathcal E_X^3\right|_\A
&{}=
\begin{bmatrix} \textcolor{blue!50}{1} \\ q \\ q \end{bmatrix}
&
\left.\mathcal E_Z^3\right|_\B
&{}=
\begin{bmatrix} 1 \\ \textcolor{blue!50}{q} \\ \overline p\,\overline q \end{bmatrix}
\end{array}
\enspace .
\end{equation}
\end{samepage}
After measuring the indicated qubit blocks, the surviving rows and columns form
$\mathcal C_1,\mathcal C_2,\mathcal C_3$ on $\A\sqcup\B\sqcup\E\sqcup\F$,
$\A\sqcup\B\sqcup\C\sqcup\D$, and $\C\sqcup\D\sqcup\E\sqcup\F$. Writing $\alpha=1+pq$,
$\beta=1+q$, and $\gamma=p+\overline q$, the corresponding stabilizer matrices are
\begin{equation}\label{eq:cIV-end}
{\setlength\arraycolsep{4pt}
\begin{array}{@{}c@{\quad}c@{\quad}c@{}}
\begin{array}{@{}c@{}}
\begin{bmatrix} H_X^1 \\\hline H_Z^1 \end{bmatrix}=
\begin{bmatrix}
\beta                     & \alpha       &                         &                         \\
                          &              & \overline\beta         & \gamma                  \\\hline
\overline\alpha           & \overline\beta & \gamma                 & p\overline\beta         \\
\overline\gamma           & \beta        &                         &
\end{bmatrix}
\end{array}
&
\begin{array}{@{}c@{}}
\begin{bmatrix} H_X^2 \\\hline H_Z^2 \end{bmatrix}=
\begin{bmatrix}
\beta                     & \alpha       &                         &                         \\
                          &              & \beta                  & \alpha                  \\\hline
\overline\gamma           & \beta        & \overline\alpha        & \overline\beta          \\
                          &              & \overline\alpha        & \overline\beta
\end{bmatrix}
\end{array}
&
\begin{array}{@{}c@{}}
\begin{bmatrix} H_X^3 \\\hline H_Z^3 \end{bmatrix}=
\begin{bmatrix}
\beta                     & \alpha       &                         &                         \\
                          &              & \overline\beta         & \gamma                  \\\hline
                          &              & \gamma                 & p\overline\beta         \\
\overline\alpha           & \overline\beta & \overline q\,\overline\alpha & \overline q\,\overline\beta
\end{bmatrix}
\end{array}
\end{array}}\enspace .
\end{equation}

\section{Numerics}\label{sec:numerics}

\subsection{Code search}\label{sec:code-search}

For numerical searches based on Constructions~II--IV, one can sample finite groups and use left-
and right-regular representations of the sampled group elements so that the required commutation
is automatic.
For Construction~I, a convenient specialization further equates the \(P\)- and \(S\)-terms and the \(Q\)- and \(R\)-terms, so that \(P=S\) and \(Q=R\).
This specialization is stricter than Criterion~C.4 of Ref.~\cite{shaw2025lowering}, which allows
the paired entries to arise as distinct coset restrictions, i.e., \(P\ne S\) or \(Q\ne R\).
More explicitly, if a two-block group algebra code with \(A=\{a_i\}\) and \(B=\{b_j\}\) satisfies
Criterion~C.4 with \(f(a_1)=f(b_1)=1\), then realizing it inside the \(P=S\), \(Q=R\)
specialization requires a bijection \(\varphi:K\to K^c\) such that \(\varphi(\varphi(g)h)=gh\)
for every \(g\in K\) and \(h\in (A\cup B)\setminus\{a_1,b_1\}\).

Tables~\ref{tab:numerical-candidates} and~\ref{tab:numerical-parameters} summarize the five
mid-cycle codes used in the circuit simulations: the \([\![288,12,18]\!]\) bivariate bicycle code
of Ref.~\cite{bravyi2024high}, simulated in Ref.~\cite{shaw2025lowering}, and one representative
from each of Constructions~I--IV.
The first table gives the code parameters, while the second provides the algebraic specifications
used to instantiate the permutation matrices, with \(n=|G|\) for rows I--IV.
For a group algebra element $\sum_i g_i\in\mathbb F_2[G]$, we write
$\mathsf L(\sum_i g_i)=\sum_i\mathsf L(g_i)$ and
$\mathsf R(\sum_i g_i)=\sum_i\mathsf R(g_i)$, where $\mathsf L(g)$ and $\mathsf R(g)$ are the
left- and right-regular representations of $g\in G$.
The ST rows of Tables~\ref{tab:numerical-candidates} and~\ref{tab:numerical-parameters} spell out
the embedding of the \([\![288,12,18]\!]\) code into Construction~I through the homomorphism
\(f=f_{xy}\) in Criterion~C.4 of Ref.~\cite{shaw2025lowering}.
Specifically, \(n=72^\ast\) in Table~\ref{tab:numerical-candidates} indicates the \((K,K^c)\)
splitting of the \(144\) group elements, and the permutation matrices in
Table~\ref{tab:numerical-parameters} are obtained by normalizing the listed translations by
\(y^2\) or \(x^2\) and then restricting to the \(K\) and \(K^c\) sectors.
For rows I--IV, the distances in Table~\ref{tab:numerical-candidates} were certified from the
specifications in Table~\ref{tab:numerical-parameters} using the exact distance solver
\texttt{pySATDist} in the \texttt{codedistance} package~\cite{webster2026distance}.

\begin{table}[p]
\caption{\justifying Code parameters and qubit connectivity degrees for the five mid-cycle codes
used in the circuit simulations. The ST row is the \([\![288,12,18]\!]\) bivariate bicycle code of
Ref.~\cite{bravyi2024high}, simulated in Ref.~\cite{shaw2025lowering}; rows I--IV are selected
from the finite-group search. For rows I--IV, \(n\) is the group size; in the ST row, \(n=72^\ast\)
is the permutation-matrix size for the Construction~I embedding.}
\label{tab:numerical-candidates}
\centering
\renewcommand{\arraystretch}{1.0}
\begin{tabular}{@{}c@{\quad}|@{\quad}c@{\quad}c@{\quad}c@{\quad}c@{\quad}c@{}}
\hline\hline
& $n$ & Connectivity & $[\![N,K,D]\!]$ & $N_j$ & $\min_j D_j$ \\
\hline
ST &
$72^\ast$ &
$5$ &
$[\![288,12,18]\!]$ &
$144$ &
$12$ \\
I &
$72$ &
$5$ &
$[\![288,16,16]\!]$ &
$144$ &
$12$ \\
II &
$65$ &
$4$ &
$[\![260,10,14]\!]$ &
$130$ &
$11$ \\
III &
$28$ &
$4$ &
$[\![224,8,16]\!]$ &
$112$ &
$12$ \\
IV &
$41$ &
$3$ &
$[\![246,4,10]\!]$ &
$164$ &
$9$ \\
\hline\hline
\end{tabular}

\vspace{0.5em}

\caption{\justifying Algebraic specifications used to instantiate the permutation matrices for
the codes in Table~\ref{tab:numerical-candidates}. In the ST row, the listed translations are
normalized by \(y^2\) or \(x^2\) and then restricted to the \(K\) and \(K^c\) sectors. For rows
I--IV, \(\mathrm{SmallGroup}\) denotes the identifier in the GAP Small Groups
Library~\cite{smallgrp}.}
\label{tab:numerical-parameters}
\small
\renewcommand{\arraystretch}{0.94}
\begin{tabular}{@{}p{\textwidth}@{}}
\hline\hline
\textbf{ST.} Ref.~\cite[Table~I, last row]{shaw2025lowering}, $(\ell,m)=(12,12)$, \(A=\{y^2,y^7,x^3\}\), \(B=\{x^2,x,y^3\}\), \(f=f_{xy}\)\\
The translations \(y^7,x^3\), normalized by \(y^2\), give \((q_1,r_1),(q_2,r_2)\) by their \((K,K^c)\) restrictions, while \(x,y^3\), normalized by \(x^2\), give \((p_1,s_1),(p_2,s_2)\); set \(Q=q_1+q_2\), \(R=r_1+r_2\), \(P=p_1+p_2\), \(S=s_1+s_2\).
\\
\hline
\textbf{I.} \(G=\mathrm{SmallGroup}(72,20)\cong(C_3:C_4)\times S_3\), \(P=S=\mathsf L(a_1+a_2)\), \(Q=R=\mathsf R(b_1+b_2)\)\\
\(G=\langle g_1,g_2,g_3;\ g_2=g_1^2,\ g_1^4=g_3^3=1,\ g_1^{-1}g_3g_1=g_3^{-1}\rangle\times\langle g_4,g_5;\ g_4^2=g_5^3=1,\ g_4g_5g_4=g_5^{-1}\rangle\)\\[-1pt]
\(a_1=g_1,\quad a_2=g_1^{-1}g_5g_3,\quad b_1=g_5g_3,\quad b_2=g_5^{-1}g_3^{-1}g_4\)
\\
\hline
\textbf{II.} \(G=\mathrm{SmallGroup}(65,1)\cong C_5\times C_{13}\), \(P=\mathsf L(a_1+a_2+a_3)\), \(q=\mathsf R(b)\)\\
\(G=\langle g_1,g_2\mid g_1^5=g_2^{13}=1,\ [g_1,g_2]=1\rangle\)\\[-1pt]
\(a_1=g_2^3g_1^2,\quad a_2=g_2^{-1},\quad a_3=g_2^{-3}g_1^2,\quad b=g_2^4\)
\\
\hline
\textbf{III.} \(G=\mathrm{SmallGroup}(28,4)\cong C_2\times C_2\times C_7\), \(P=\mathsf L(a_1+a_2)\), \(Q=\mathsf L(b_1+b_2)\), \(r=\mathsf R(c)\)\\
\(G=\langle g_1,g_2,g_3\mid g_1^2=g_2^2=g_3^7=1,\ [g_i,g_j]=1\rangle\)\\[-1pt]
\(a_1=g_3^{-3}g_2,\quad a_2=g_3^{-3}g_2g_1,\quad b_1=g_3^{-2}g_2g_1,\quad b_2=g_3^2g_1,\quad c=g_3^{-1}g_2g_1\)
\\
\hline
\textbf{IV.} \(G=\mathrm{SmallGroup}(41,1)\cong C_{41}=\langle g\mid g^{41}=1\rangle\), \(p=\mathsf L(g)\), \(q=\mathsf R(g^4)\)
\\
\hline\hline
\end{tabular}
\end{table}

\subsection{Circuit simulation}\label{sec:circuit-simulation}

\paragraph{Operations and observables.}\phantomsection\label{sec:observables}
For each of the five mid-cycle codes in Tables~\ref{tab:numerical-candidates}
and~\ref{tab:numerical-parameters}, we instantiate the corresponding morphing circuit and simulate
four cycles with circuit-level depolarizing noise using Stim~\cite{gidney2021stim}.
A cycle starts and ends at $\mathcal C_1$, as in Fig.~\ref{fig:bowtie}.
For $J=3$, the same convention gives \(R_1,\ F_1^\dagger,\ F_2,\ M_2,\ R_2,\ F_2^\dagger,\ F_3,\ M_3,\ R_3,\ F_3^\dagger,\ F_1,\ M_1\).
Here, $M_j$ denotes the measurements of the contracted stabilizers in round $j$, and $R_j$ denotes
the corresponding resets of the same qubits in the same bases.
Within each $F_j$, the $\CX$ layers in each polynomial are applied in the arbitrary fixed order
listed in Table~\ref{tab:numerical-parameters}; in $F_j^\dagger$, the corresponding layers are
applied in the inverse order.
Although this ordering freedom can affect the resulting circuit distance, we do not optimize over it.
At noise rate \(p\), each operation is independently corrupted: $\MX$ and $\MZ$ outcomes are
flipped, $\RX$ ($\RZ$) is followed by \(Z\) (\(X\)) error, and $\CX$ is followed by one of the
15 nonidentity Pauli errors chosen uniformly.

Following the magic boundary convention in Ref.~\cite{gidney2025yoked}, we introduce \(K\)
reference qubits and prepare a noiseless Bell state stabilized by \(X_i\otimes \overline X_i\)
and \(Z_i\otimes \overline Z_i\), \(i\in[K]\), where
\(\{\overline X_i,\overline Z_i\}_{i\in[K]}\) is a symplectic logical basis of \(\mathcal C_1\).
After the four noisy cycles, we noiselessly measure the checks of \(\mathcal C_1\), and then
noiselessly measure \(X_i\otimes \overline X'_i\) and \(Z_i\otimes \overline Z'_i\), where
\(\{\overline X'_i,\overline Z'_i\}_{i\in[K]}\) is the evolved symplectic logical basis of
\(\mathcal C_1\).
The observables thus consist of the measurement records accumulated during the four cycles together with these final logical-basis measurements.
The evolution of the logical basis through the gates and measurements of the morphing circuit has
no closed form; in the implementation, it is obtained by binary matrix operations following
Eq.~\eqref{eq:cxrule}.

\paragraph{Detector construction.}\phantomsection\label{sec:detectors}
Detectors away from the initial and final time boundaries have closed forms, derived from
Eqs.~\eqref{eq:cI-evolved}, \eqref{eq:cII-evolved}, \eqref{eq:cIII-evolved},
and~\eqref{eq:cIV-evolved} and listed in Table~\ref{tab:bulk-detectors}.
For a monomial $p$, $p(i)$ denotes the column index $j$ such that $p_{ij}=1$.
Within each cycle, each qubit is measured at most once in each basis, and every stabilizer of the mid-cycle code is measured exactly once.
We write $[\B,i]_X^\tau$ for the outcome of measuring qubit $i\in[n]$ in qubit block $\B$ in
the $X$-basis during cycle $\tau$, and similarly for $Z$-basis measurements and other qubit blocks.
We write $\{\mathsf c_1,i\}_X^\tau$ for the detector associated with the reset of the $X$-type
stabilizer $i\in[n]$ in stabilizer block $\mathsf c_1$ during cycle $\tau$, and similarly for
$Z$-type stabilizers and other stabilizer blocks.
All sums of outcomes are over $\mathbb F_2$.

\begin{center}
\begin{minipage}{\textwidth}
\captionsetup{type=table,hypcap=false}
\caption{\justifying Bulk detectors for the four constructions.}
\label{tab:bulk-detectors}
\footnotesize
\renewcommand{\arraystretch}{1.0}
\begin{tabular}{@{}p{0.49\textwidth}@{\hspace{0.01\textwidth}}p{0.49\textwidth}@{}}
\hline\hline
\multicolumn{2}{@{}l@{}}{\textbf{I.} $P=\sum_\alpha p_\alpha$, $Q=\sum_\beta q_\beta$, $R=\sum_\gamma r_\gamma$, $S=\sum_\delta s_\delta$}\\
\(\begin{aligned}[t]
\{\mathsf c_1,i\}_X^\tau &= \sum_\alpha [\B,p_\alpha(i)]_X^\tau+[\D,i]_X^\tau\\
\{\mathsf c_4,i\}_Z^\tau &= \sum_\beta [\D,\overline q_\beta(i)]_Z^\tau+[\B,i]_Z^\tau
\end{aligned}\)
&
\(\begin{aligned}[t]
\{\mathsf c_2,i\}_X^\tau &= \sum_\delta [\D,s_\delta(i)]_X^\tau+[\B,i]_X^{\tau+1}\\
\{\mathsf c_3,i\}_Z^\tau &= \sum_\gamma [\B,\overline r_\gamma(i)]_Z^\tau+[\D,i]_Z^{\tau+1}
\end{aligned}\)\\
\hline
\multicolumn{2}{@{}l@{}}{\textbf{II.} $P=\sum_\alpha p_\alpha$}\\
\(\begin{aligned}[t]
\{\mathsf c_1,i\}_X^\tau &= \sum_\alpha [\B,p_\alpha(i)]_X^\tau+[\C,i]_X^\tau\\
\{\mathsf c_2,i\}_Z^\tau &= \sum_\alpha [\C,\overline p_\alpha(i)]_Z^\tau+[\B,i]_Z^\tau
\end{aligned}\)
&
\(\begin{aligned}[t]
\{\mathsf c_2,i\}_X^\tau &= \sum_\alpha [\C,\overline p_\alpha(i)]_X^\tau+[\B,i]_X^{\tau+1}\\
\{\mathsf c_1,i\}_Z^\tau &= \sum_\alpha [\B,p_\alpha(i)]_Z^\tau+[\C,i]_Z^{\tau+1}
\end{aligned}\)\\
\hline
\multicolumn{2}{@{}l@{}}{\textbf{III.} $P=\sum_\alpha p_\alpha$, $Q=\sum_\beta q_\beta$}\\
\(\begin{aligned}[t]
\{\mathsf c_1,i\}_X^\tau &= \sum_\beta [\H,r\overline q_\beta(i)]_X^\tau+[\A,i]_X^\tau+[\B,i]_X^\tau\\
\{\mathsf c_1,i\}_Z^\tau &= \sum_\beta [\H,r\overline q_\beta(i)]_Z^\tau+[\A,i]_Z^\tau+[\B,i]_Z^{\tau+1}
\end{aligned}\)
&
\(\begin{aligned}[t]
\{\mathsf c_3,i\}_X^\tau &= \sum_\beta [\B,\overline r q_\beta(i)]_X^\tau+[\G,i]_X^\tau+[\H,i]_X^{\tau+1}\\
\{\mathsf c_3,i\}_Z^\tau &= \sum_\beta [\B,\overline r q_\beta(i)]_Z^\tau+[\G,i]_Z^\tau+[\H,i]_Z^\tau
\end{aligned}\)\\
\multicolumn{2}{@{}l@{}}{\(\begin{aligned}[t]
\{\mathsf c_2,i\}_X^\tau &= \sum_\alpha [\A,p_\alpha(i)]_X^\tau+\sum_{\beta,\beta'} [\H,rq_\beta\overline q_{\beta'}(i)]_X^\tau+[\H,r(i)]_X^\tau+[\G,r(i)]_X^\tau\\
\{\mathsf c_2,i\}_Z^\tau &= \sum_\alpha [\A,p_\alpha(i)]_Z^\tau+\sum_{\beta,\beta'} [\H,rq_\beta\overline q_{\beta'}(i)]_Z^\tau+[\H,r(i)]_Z^\tau+[\G,r(i)]_Z^{\tau+1}\\
\{\mathsf c_4,i\}_X^\tau &= \sum_\alpha [\G,\overline p_\alpha(i)]_X^\tau+\sum_{\beta,\beta'} [\B,\overline r\,\overline q_\beta q_{\beta'}(i)]_X^\tau+[\B,\overline r(i)]_X^\tau+[\A,\overline r(i)]_X^{\tau+1}\\
\{\mathsf c_4,i\}_Z^\tau &= \sum_\alpha [\G,\overline p_\alpha(i)]_Z^\tau+\sum_{\beta,\beta'} [\B,\overline r\,\overline q_\beta q_{\beta'}(i)]_Z^\tau+[\B,\overline r(i)]_Z^\tau+[\A,\overline r(i)]_Z^\tau
\end{aligned}\)}\\
\hline
\multicolumn{2}{@{}l@{}}{\textbf{IV.}}\\
\(\begin{aligned}[t]
\{\mathsf c_1,i\}_X^\tau &= [\C,pq(i)]_X^\tau+[\E,p(i)]_X^{\tau+1}+[\A,i]_X^{\tau+1}\\
\{\mathsf c_2,i\}_X^\tau &= [\E,pq(i)]_X^\tau+[\A,q(i)]_X^\tau+[\C,i]_X^\tau\\
\{\mathsf c_3,i\}_X^\tau &= [\A,q(i)]_X^\tau+[\C,i]_X^\tau+[\E,\overline q(i)]_X^{\tau+1}
\end{aligned}\)
&
\(\begin{aligned}[t]
\{\mathsf c_1,i\}_Z^\tau &= [\B,i]_Z^\tau+[\D,\overline q(i)]_Z^\tau+[\F,p(i)]_Z^{\tau+1}\\
\{\mathsf c_2,i\}_Z^\tau &= [\D,i]_Z^\tau+[\F,\overline q(i)]_Z^{\tau+1}+[\B,q(i)]_Z^{\tau+1}\\
\{\mathsf c_3,i\}_Z^\tau &= [\F,\overline q(i)]_Z^\tau+[\B,\overline p\,\overline q(i)]_Z^\tau+[\D,i]_Z^\tau
\end{aligned}\)\\
\hline\hline
\end{tabular}
\end{minipage}
\end{center}

\paragraph{Sampling and decoding.}

The decoder is the batched GPU implementation of Relay-BP~\cite{muller2025improved} in NVIDIA CUDA-Q QEC~\cite{cudaq-qec}.
We draw detector flips and observable flips from the full Stim detector error model, containing both \(X\)- and \(Z\)-type sectors.
Each sampled shot is then split into its \(X\)- and \(Z\)-type sectors, decoded separately, and
declared a failure if either sector predicts an incorrect observable flip.
We report the empirical block logical error rate per cycle, namely, the logical failure probability per shot divided by four.

\begin{figure}[t]
\centering
\includegraphics[width=0.82\linewidth]{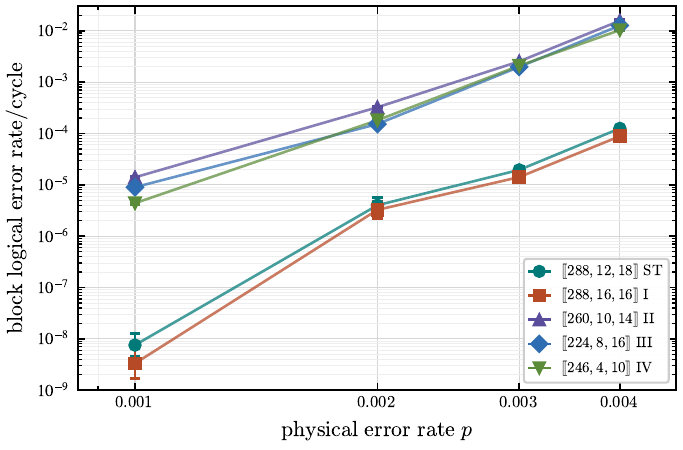}
\caption{\justifying Circuit-level block logical error rates per cycle for the Shaw--Terhal two-block group algebra circuit (ST) and Constructions~I--IV.
Error bars are 95\% Wilson score intervals for the logical failure rate, rescaled per cycle.}
\label{fig:logical-error-preview}
\end{figure}

For each circuit and noise rate \(p=0.001,\ldots,0.004\), we first run a preliminary grid sweep over \(\gamma_{\rm low}\) and \(\gamma_{\rm high}\).
We choose the \(\gamma\) interval with the smallest empirical logical failure rate in this sweep,
and keep it fixed for the simulations reported in Fig.~\ref{fig:logical-error-preview}.
All reported decoder runs use \(\gamma_0=0.125\), \(\texttt{max\_iterations}=60\),
\(\texttt{pre\_iter}=80\), \(\texttt{num\_sets}=100\), \(\texttt{stop\_nconv}=5\), and no OSD.
The resulting circuit-level logical error rates are shown in Fig.~\ref{fig:logical-error-preview},
and the selected \(\gamma\) intervals are listed in Table~\ref{tab:gamma-windows}.

\begin{table}[t]
\caption{\justifying Relay-BP \(\gamma\) intervals \([\gamma_{\rm low},\gamma_{\rm high}]\)
used for the simulations reported in Fig.~\ref{fig:logical-error-preview}; columns are labeled
by the physical error rate \(p\).}
\label{tab:gamma-windows}
\centering
\footnotesize
\setlength{\tabcolsep}{3.5pt}
\begin{tabular}{@{}ccccc@{}}
\hline\hline
& \(0.001\) & \(0.002\) & \(0.003\) & \(0.004\)\\
\hline
ST  & \([-0.20,0.75]\)     & \([-0.03,0.57]\) & \([-0.08,0.58]\)  & \([-0.145,0.565]\)\\
I   & \([-0.20,1.00]\)     & \([-0.11,0.45]\) & \([-0.12,0.50]\)  & \([-0.148,0.712]\)\\
II  & \([-0.21,0.55]\)     & \([-0.11,0.45]\) & \([-0.096,0.564]\)& \([-0.040,0.620]\)\\
III & \([-0.18,0.68]\)     & \([-0.14,0.72]\) & \([-0.047,0.563]\)& \([-0.066,0.694]\)\\
IV  & \([-0.19,0.77]\)     & \([-0.14,0.72]\) & \([-0.070,0.650]\)& \([-0.073,0.637]\)\\
\hline\hline
\end{tabular}
\end{table}

\section{Outlook}

Many questions remain open.
First, for the constructions in Table~\ref{tab:constructions}, the relation between the mid-cycle
distance $D$ and the end-cycle distances $D_j$ is understood only through the depth-dependent lower
bounds of Refs.~\cite{shaw2025lowering,shaw2026optimising}.
Most instances encountered during our numerical search have $D>D_j$, but a few have $D<D_j$.
It would be useful to find sharper constraints on $D_j$ and to understand how $D_j$ changes under
different choices of pivots, such as the choices leading to Eqs.~\eqref{eq:cII-end}
and~\eqref{eq:cII-pivot-end}.

Second, morphing circuits based on subsystem codes give another possible extension.
The diamond circuits for surface codes introduced by Debroy~\cite{debroy2025diamond} can be
written in block-algebra form, but with only binomial block entries for the mid-cycle stabilizer
code, leaving no monomial pivots for the Gaussian eliminations used in the present constructions.

Third, the block-algebra formalism might provide a useful language for designing logical operations
in morphing circuits~\cite{shaw2025lowering,shaw2026optimising,chen2024transversal}.

Fourth, the constructions in Table~\ref{tab:constructions} have zero encoding rate.
An ambitious direction is to design morphing circuits based on high-rate codes, such as Kasai's construction~\cite{kasai2026breaking}.

\begin{acknowledgments}
The author is grateful to Yifan Hong, Krysta Svore, Vadym Kliuchnikov, Adam Holmes, and Muyuan Li for helpful discussions.
Large language models were used to assist with numerical scripting and manuscript editing;
the author checked the content of this work and is responsible for it.
\end{acknowledgments}

\bibliographystyle{q}
\bibliography{q}
\end{document}